\documentstyle[prl,aps,floats,epsf]{revtex}\def\narrowtext{}\tighten\twocolumn
\input epsf.sty
\begin{document}
\draft

\title{Cu Nuclear Quadrupole Resonance Study of 
Site-Disorder and Chemical Pressure Effects on 
Y(Ba$_{1-x}$Sr$_x$)$_2$Cu$_4$O$_8$}
\author{Y. Itoh$^{1,2}$, S. Adachi$^1$, T. Machi$^1$, and N. Koshizuka$^1$}

\address{$^1$Superconductivity Research Laboratory,
 International Superconductivity Technology Center,\\
1-10-13 Shinonome, Koto-ku Tokyo 135-0062, Japan \\
$^2$Japan Science and Technology Corporation, Japan}


\address{\begin{minipage}[t]{6.0in} %
\begin{abstract}%
We report a zero-field Cu nuclear quadrupole resonance (NQR) study on the effects of
nonmagnetic Sr substitution for high-$T_c$ superconductors, Y(Ba$_{1-x}$Sr$_x$)$_2$Cu$_4$O$_8$
($T_c$=82-80 K for $x$=0-0.4), using a spin-echo technique. The site-disordering and chemical
pressure effects associated with doping Sr were observed in the broadened, shifted Cu NQR
spectra. Nevertheless, the site disorder did not significantly affect the homogeneity of Cu
electron spin dynamics, in contrast to the in-plane impurity. The peak shift of Cu NQR
spectrum due to Sr was different between the chain- and the plane-Cu sites, more remarkably
than those under a hydrostatic physical pressure, suggesting anisotropic or nonuniform local
structural strains. The small decrease of $T_c$ due to Sr can be traced back to either a
cancellation effect on $T_c$ between the disorder and the pressure, or an
anisotropic or nonuniform chemical pressure effect on $T_c$.   
\typeout{polish abstract} %
\end{abstract}
\end{minipage}} %

\maketitle %
\narrowtext

The role of imperfections in quantum many-body systems has attracted strong
attention~\cite{Anderson,Sachdev}. In high-$T_c$ cuprate superconductors, an additional
impurity potential is introduced through the in-plane or out-of-plane element substitution
for the CuO$_2$ plane. There is a belief that site disorder does not cause serious damage
to the electronic states of the CuO$_2$ plane. However, the reason why chemical pressure
effect on $T_c$ due to the element substitution is not the same as the physical (external)
pressure one has been frequently ascribed to the randomness effect. How the site disorder
changes the electronic states has been a problem.    

Here, we focus on the divalent Sr$^{2+}$-doped high-$T_c$ superconductors,
Y(Ba$_{1-x}$Sr$_x$)$_2$Cu$_4$O$_8$. Since the stoichiometric and naturally underdoped
YBa$_2$Cu$_4$O$_8$ (Y124) with $T_c$=82 K has the exceptionally tight oxygen content, the
Sr-doped Y124 is a suitable system to study the site-disordering effect on the electronic
states. To be exact, one can expect two effects of Sr substitution for Ba in Y124 without
change of oxygen content; (1) the chemical pressure and (2) the crystalline potential
disorder. The substitution of Sr$^{2+}$ ions causes no additional local (formal) charge in
Ba$^{2+}$O$^{2-}$ layers. However, the size of Sr$^{2+}$ ion is smaller than that of
Ba$^{2+}$ by about 10 $\%$ ~\cite{Shannon}, so that the substituted Sr ions make local
crystalline strains on the BaO layers. Thus, the Sr doping introduces crystalline potential
disorder and chemical pressure to the lattice. Actually, the lattice constants shrink with
doping Sr~\cite{Wada1}, being in parallel to the physical pressure effect~\cite{Yamada}. 

The effect of hydrostatic physical pressure on the spin dynamics is similar to 
the carrier doping effect~\cite{Machi1,Machi2,Machi3}: The physical pressure effects on
$T_c$ and on the pseudo spin-gap temperature $T_s$ agree with the carrier doping effects,
i.e. with applying the physical pressure to Y124 ($T_c$=82 K and $T_s\sim$160 K at an ambient
pressure $P$=0.1 MPa), $T_c$ increases~\cite{Kaldis} but $T_s$ decreases~\cite{Machi3}, as
well as with doping Ca~\cite{Miyatake1,Machi1}. Here, the pseudo spin-gap temperature
$T_s$ is defined as the maximum temperature of the planar
$^{63}$Cu(2) nuclear spin-lattice relaxation rate $^{63}(1/T_1T)$ ~\cite{Yasuoka}, which is
the wave-vector averaged dynamical spin susceptibility at a nuclear quadrupole resonance
 (NQR)/NMR frequency via a nuclear-electron coupling constant~\cite{Moriya}. It is reported 
that the uniaxial
pressure also increases $T_c$ except along the $c$-axis ($dT_c/dP_c\sim$0, $P_c$ is a
pressure along the $c$-axis) from the thermal expansion coefficients of the lattice
constants for powder Y124 ~\cite{Meingast}.

The Sr doping, however, scarcely increases $T_c$ for Y124 ~\cite{Wada1}, or remarkably
decreases $T_c$ for YBa$_2$Cu$_3$O$_{7-\delta}$ ~\cite{Wada2,Oka,Karen,Licci,Cao}, although
the unit cell volume shrinks (the unit cell volume of Y124 with Sr of $x$=0.3 corresponds to
that under an external hydrostatic pressure of about 2.4 GPa~\cite{Ishigaki}). Hence, one
can suspect randomness effect on the actual
$T_c$. Nonmagnetic impurity Zn doping into the CuO$_2$ plane significantly decreases $T_c$
in the hole-doped high-$T_c$ cuprates~\cite{Xiao}. However, there is a crucial
difference between Zn$^{2+}$ in the CuO$_2$ plane and Sr$^{2+}$ in the BaO layer: The
Zn$^{2+}$ ion acts as a spin vacancy in the CuO$_2$ plane with site disorder, whereas the
Sr$^{2+}$ ion acts only as a site disorder in the CuO$_2$ plane without any vacancies. In
terms of a single-band Hubbard model for the CuO$_2$ plane, the site disorder is descibed
by the randomness or the modulation of transfer matrix elements $t_{ij}$ ($i$ and $j$ are
the labels of the nearest-neighbor Cu sites) in the kinetic energy of the conducting
carriers $\sum_{i,j} t_{ij}c_i^{\dagger}c_j$ ($c_i^{\dagger}$ and $c_j$ are the creation 
and the annihilation operators of an electron), which may cause a distribution of the local
 density of states of electrons.

In the charge-transfer model from the CuO chain to the CuO$_2$ plane, a charge distribution
between the chain Cu(1) and the plane Cu(2) plays an important role in changing
$T_c$ ~\cite{Tokura,Jorgensen,Yamada,Licci}. From the structural analysis,
however, the bond valence sums at Cu(1) and at Cu(2) are estimated to increase with doping
Sr for Y124~\cite{Ishigaki}. From the theoretical analysis based on a band theory, an
important role of an internal strain of the CuO$_2$ plane is emphasized~\cite{Pickett}. Cu 
NQR spectrum can provide information on the charge
distribution at each Cu site.

In this paper, we report on Cu NQR measurements for Y(Ba$_{1-x}$Sr$_x$)$_2$Cu$_4$O$_8$. We
observed the site-disordering and chemical pressure effects in the broadened, shifted Cu
NQR spectra. We found that the site disorder does not significantly affect the homogeneity
of the electron spin dynamics of the CuO$_2$ plane, sharply in contrast to that due to the
in-plane impurities. 

The powdered samples were synthesized by a solid-state-reaction method with a
hot-isostatic-pressing apparatus. From measurements of the $dc$ magnetization with a SQUID
magnetometer, the superconducting transition was observed at $T_c$=82, $\sim$81,
$\sim$80.5, and $\sim$80 K for $x$=0,00, 0.10, 0.20, and 0.40, respectively. While the
previous study reported somewhat increase in $T_c$ with increasing Sr content~\cite{Wada1}, our samples
behaved conversely. Our samples were prepared through rather long-time heat-treatment under
high oxygen partial pressure. We believe that the samples were purely single-phasic and free
from structural defects, e.g. Cu-O single-chain. Synthesis and characterization of the
samples will be reported elsewhere in details~\cite{Adachi}.

\begin{figure}
\epsfxsize=2.8in
\epsfbox{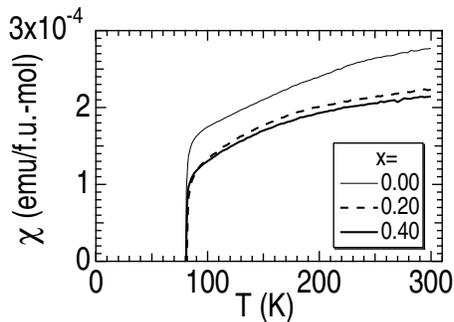}
\vspace{-0.3cm}
\caption{
The temperature and Sr-doping dependence of the static uniform susceptibility $\chi$ of
Y(Ba$_{1-x}$Sr$_x$)$_2$Cu$_4$O$_8$ ($x$=0, 0.20, and 0.40). In contrast to Zn substitution for
Cu, no Curie term is induced by Sr substitution for Ba.
}
\label{X}
\end{figure}

NQR measurements were carried out by a coherent-type pulsed spectrometer.
The Cu NQR spectra were measured by integration of the Cu nuclear spin-echoes as the
frequency is changed. The 63Cu(2) nuclear spin-lattice relaxation curves
$p(t)\equiv1-M(t)/M(\infty)$ were measured by an inversion recovery technique, where the
$^{63}$Cu nuclear spin-echo intensity $M(t)$ was recorded as a function of time interval $t$
after an inversion $\pi$ pulse, in a $\pi-t-\pi/2-\pi$-echo sequence. The recovery curve $p(t)$
was analyzed by $p_{LD}(t)=p_{LD}(0)$exp$[-(3t/T_1)_{HOST}-\sqrt{3t/\tau_1}]$, where
$p_{LD}(0)$, $(T_1)_{HOST}$ and $\tau_1$ are the fitting parameters. $p_{LD}(0)$ is a
fraction of an initially inverted magnetization, $(T_1)_{HOST}$ is the nuclear spin-lattice
relaxation time due to the host Cu electron spin fluctuations, and $\tau_1$ is an
impurity-induced nuclear spin-lattice relaxation time, which is originally called a
longitudinal direct dipole relaxation time~\cite{McHenry}. We regard $\tau_1$ as just the
measure of deviation from the single exponential function. An alternative analysis by a
two-exponential function of $p_S$exp$(-t/T_{1S})+p_L$exp$(-t/T_{1L})$ ($p_S$, $T_{1S}$, $p_L$,
and $T_{1L}(>T_{1S}$) are the fitting parameters) is possible, but it remains our conclusions
unchanged. We emphasize that the present result is independent of the details of analysis. 

Figure 1 shows the uniform spin susceptibility $\chi$ for the powdered samples, which were
measured using a SQUID magnetometer under an external magnetic field of 1 $T$. Obviously, no
Curie term is observed, sharply in contrast to the Zn doped Y124~\cite{Miyatake2}. The site
disorder does not induce a local moment. The small decrease of $\chi$ with doping Sr cannot
be accounted for by the  diamagnetic susceptibility $\chi_{core}$ of core electrons
of Sr$^{2+}$ ions, because of the estimated $\chi_{core}$=-(2.16-2.02)x10$^{-4}$
(emu/f.u.-mol) for
$x$=0-0.40 [$\chi_{core}$=34$x$-216 (10$^{-6}$ emu/f.u.-mol) for $x$=0-2]. The origin of
the decrease is not clear at present. 

\begin{figure}
\epsfxsize=2.5in
\epsfbox{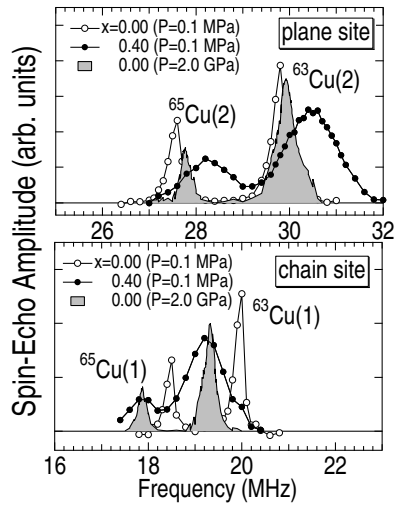}
\vspace{0cm}
\caption{
The $^{63, 65}$Cu NQR frequency spectra of the plane Cu(2) (a) and the chain Cu(1) (b) for
Y(Ba$_{1-x}$Sr$_x$)$_2$Cu$_4$O$_8$ of $x$=0.00 (open circles) and $x$=0.40 (closed circles) at
an ambient pressure ($P$=0.1 MPa). For comparison, the Cu NQR spectra (shaded area) of pure
Y124 under a hydrostatic physical pressure ($P$=2.0 GPa) are also reproduced from Ref. [7].
}
\label{NQR}
\end{figure} 

Figures 2 shows $^{63, 65}$Cu NQR spectra of the plane Cu(2) (a) and of the chain Cu(1) (b)
for $x$=0.00 (open circles) and $x$=0.40 (closed circles) at $T$=4.2 K. For comparison, the
shaded $^{63, 65}$Cu NQR spectra for a pure Y124 under a hydrostatic physical pressure of
$P$=2.0 GPa are reproduced from Ref.~\cite{Machi2}. With doping Sr, the Cu(2) NQR spectra
are shifted to higher frequencies, whereas the Cu(1) NQR spectra are shifted to lower
frequencies. Since the directions of these shifts are in parallel to those under the
hydrostatic pressure, a charge transfer from the chain to the plane may occur with doping
Sr, similarly to the external pressure effect~\cite{Zimmermann,Machi2}. However, one should
note that the degree of shifts of the Cu(2) NQR spectra is quite different from that of
Cu(1) between the Sr-doping and the hydrostatic pressure effects~\cite{Zimmermann,Machi2}.
This difference indicates that compression due to the internal pressure of Sr is different
from that due to the hydrostatic pressing. A local strain model as in
Ref.~\cite{Pickett} may account for the shifts of the Cu NQR spectra. 

NQR frequency indicates the deviation of charge distribution from cubic symmetry around the
nuclear site, which is quite sensitive to the local crystal structure. Both the linewidths
of Cu(1) and Cu(2) are broadened by Sr doping, indicating the enhancement of randomness of
the crystalline potential due to Sr in BaO layer. For such a broadened spectrum, one would
expect an inhomogeneous local density of electron states and an inhomogeneous electron spin
dynamics.   

\begin{figure}
\epsfxsize=2.8in
\epsfbox{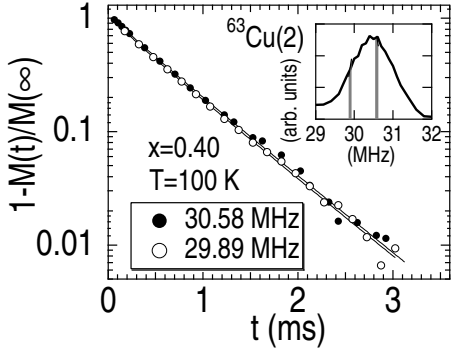}
\vspace{-0.3cm}
\caption{
The frequency dependence of the recovery curve of the $^{63}$Cu(2) nuclear spin-echo $M(t)$
at 31 MHz (closed circles) and at 29 MHz (open circles) for $x$=0.40 at 100 K. The solid curves are the
least-squares fitting results using eq. $p_{LD}(t)$ (see the text). The inset
shows the $^{63}$Cu(2) NQR spectrum and the corresponding hot spot regions (shaded area)
excited by the rf pulse in the measurement of the recovery curve.   
}
\label{Recovery}
\end{figure}

Figure 3 shows a semi-logarithmic plot of the $^{63}$Cu(2) nuclear spin-echo recovery curves
$p(t)$ at 30.89 MHz and at 29.89 MHz in the broadened spectrum of $x$=0.40 at $T$=100 K. The
inset figure shows the broadened $^{63}$Cu(2) NQR spectrum, where the shaded areas represent
the hot spot regions excited by the first $\pi$/2 pulse at the respective frequencies for
the measurement of the recovery curves. The excited region is estimated from the pulse
strength $\nu_1$ of the first $\pi$/2 pulse, e. g. $\nu_1\sim$63 kHz from the relation of
$2\pi\nu_1t_w=\pi$/2 using the time width of the first $\pi$/2 pulse, $t_w\sim~$4 $\mu$s. In
contrast to our naive expectation, the recovery curves are nearly the same single
exponential function, being independent of the frequency. The lattice imperfections broaden
the static NQR spectra, nevertheless it does not affect the homogeneity of the Cu spin
dynamics in the CuO$_2$ plane. That is, the homogeneous spin dynamics is observed in the
inhomogeneously broadened Cu NQR spectrum. 
 
\begin{figure}
\epsfxsize=2.5in
\epsfbox{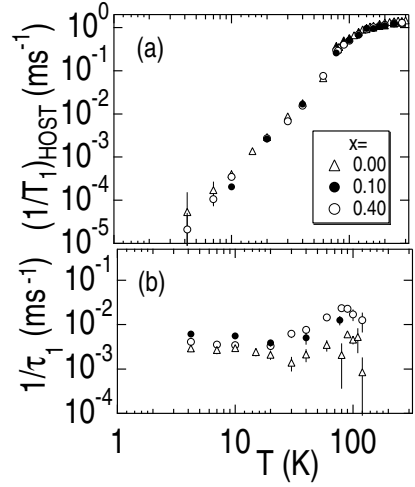}
\vspace{-0.3cm}
\caption{
The temperature and Sr-doping dependence of (1/$T_1T)_{HOST}$ (a) and 1/$\tau_1$ (b)
estimated from the $^{63}$Cu(2) nuclear spin-echo recovery curves. 
}
\label{LogW1}
\end{figure} 

Figure 4 shows the Sr-doping dependence of the $^{63}$Cu(2) nuclear spin-lattice relaxation
 rate (1/$T_1)_{HOST}$ and of the extra relaxation rate 1/$\tau_1$ in a log-log plot. The magnitude
of (1/$T_1)_{HOST}$ for $x>$0 is slightly smaller than that for $x$=0 over a temperature range of
4.2-300 K. The smaller (1/$T_1)_{HOST}$ than that for $x$=0 was also observed for the chain
$^{63}$Cu(1) (not shown here). With doping Sr, the extra relaxation rate 1/$\tau_1$ is
slightly increased. Although the small increase of 1/$\tau_1$ is only an indication of the
site disorder effect on the homogeneity of Cu electron spin dynamics, the increased
1/$\tau_1$ for $x$=0.40 is one or two orders of magnitude smaller than that for 1-2 $\%$ Zn-or
Ni-doped Y124~\cite{Itoh1,Itoh2,Itoh3,Itoh4}. Since there is no Curie term in the uniform
susceptibility in Fig. 1, the physical origin of $\tau_1$ must not be paramagnetic
impurities. The overall feature of Cu nuclear spin-lattice relaxation is not changed so
much by the Sr doping. The host Cu spin dynamics is robust and still unique. The robust
electronic state for the site disorder may be a remarkable characteristic of the
two-dimensional electron system confined within a CuO$_2$ plane, which may indicate a
"quantum protectorate"~\cite{Anderson}. 

Figure 5 shows the Sr-doping effect on the pseudo spin-gap behavior of (1/$T_1T)_{HOST}$.
The pseudo spin-gap temperature $T_s$ slightly decreases with doping Sr, similarly to the
physical pressure effect~\cite{Machi3}. The slight suppression in the magnitude of (1/$T_1T)_{HOST}$
with doping Sr is similar to that with the in-plane Zn doping~\cite{Itoh2,Itoh4}. For
the high-$T_c$ cuprates, the Cu(2) 1/$T_1T$ is approximately expressed by an antiferromagnetic
spin susceptibility. The site disorder suppresses slightly the host antiferromagnetic spin
susceptibility. Thus, we found that with doping Sr, both the inhomogenization in the Cu
spin dynamics and the suppression of the host Cu antiferromagneic spin correlation are
small as well as the small decrease of $T_c$.  

\begin{figure}
\epsfxsize=2.5in
\epsfbox{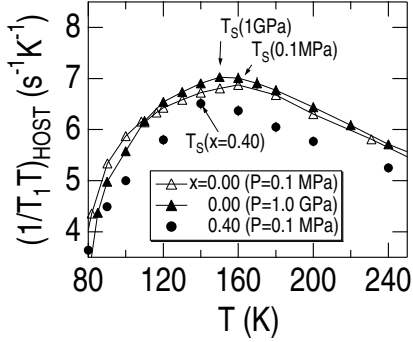}
\vspace{-0.0cm}
\caption{
The Sr-doping effect (from Fig. 4) and the hydrostatic pressure effect (reproduced from Ref.
[8]) on (1/$T_1T)_{HOST}$ around the pseudo spin-gap temperature $T_s$. The solid curves are
guides for the eye. 
}
\label{LinW1}
\end{figure}

From these results, one can infer two possibilities for the Sr-doping effect on the spin
dynamics and $T_c$. One is the cancellation mechanism between the disorder and
pressure effects. The original effect of the site disorder may be so large as to induce a
large 1/$\tau_1$ and a large decrease of $T_c$. But, the Sr chemical pressure may increase
the carrier density, as the hydrostatic physical pressure, which decreases 1/$\tau_1$ and
raises $T_c$. As a result, the Sr doping may retain a slight decrease of $T_c$ and less
inhomogeneity in the spin dynamics. The other is the anisotropic or nonuniform chemical
pressure effect on $T_c$ more than what would be expected from the thermal
expansion~\cite{Meingast}. The nonuniform compression between the plane and the
chain causes the large difference in the shifts of the Cu NQR spectra between the plane and
the chain. The cancellation among the competing anisotropic local pressure effects may
reduce the enhancement effect on $T_c$. Here, we assume that the site disorder has
essentially a weak effect on the Cu electron spin dynamics. However, it is hard to
investigate the uniaxial external pressure effects of a few GPa in polycrystalline
ceramics. A further discussion on the completely exclusive effects of the chemical pressure
and site disorder is beyond the present study. 

In conclusion, we found that the effect of the site disorder due to Sr substitution on the
Cu(2) electronic states of Y124 is quite different from that due to the in-plane impurities
such as Zn and Ni. The site disorder due to Sr does not induce local moments nor any
significant inhomogeneous relaxation process, but causes chemical pressure shifts and
broadening in the Cu NQR spectra. Further investigations, e.g. physical pressure and/or
carrier doping dependence of the Sr doping effect, will have to be conducted to separate the
roles of the chemical pressure and the site disorder.   
 
This work was supported by New Energy and Industrial Technology Development Organization
(NEDO) as Collaborative Research and Development of Fundamental Technologies for
Superconductivity Applications.

\end{document}